\documentclass[twocolumn]{aastex701}
\usepackage{float}
\usepackage{amsmath}   
\usepackage{graphicx}
\usepackage{siunitx}

\begin{document}

\title{The Apparent Asymmetric Outflows of TeV Particles from Pulsar Winds}

\author{Hongyu Pu}
\affiliation{School of Physical Science and Technology, Southwest Jiaotong University, Chengdu 610031, People's Republic of China}
\email{632606346@qq.com}

\author{Siming Liu}
\affiliation{School of Physical Science and Technology, Southwest Jiaotong University, Chengdu 610031, People's Republic of China}
\email{liusm@swjtu.edu.cn}

\begin{abstract}
Observations of X-ray filaments attached to a couple of powerful pulsars suggest escape of TeV electrons and/or positrons (e$^{\pm}$) from pulsar bow shocks into surrounding large scale magnetic fields. These filaments are usually asymmetric with very weak emission from the other side of the main filaments, and no significant spectral variation has been detected across these filaments, implying inefficient energy loss of emitting particles. We develop a Monte Carlo code to simulate particle transport in a large scale magnetic field and apply the model to PSR B2224+4415 (Guitar). It is shown that, with an injection power of a few tens of percent of the pulsar spin down luminosity, TeV e$^{\pm}$ can explain the observed filament properties with a scattering mean free path along the magnetic field comparable to the length of the observed filament. The model predicts a dim diffuse symmetric X-ray background aligned with the filament on a larger scale, whose flux is proportional to the X-ray emitting e$^{\pm}$ energy loss time for a stable e$^{\pm}$ injection power comparable to the luminosity of this diffuse background. Observations with a large field of view and good sensitivity should be able to detect such a component. 


\end{abstract}


\section{Introduction}

X-ray filaments are ribbon features attached to powerful pulsars \citep{2024ApJ...976....4D, 2022ApJ...939...70D}. 
They are produced by TeV e$^{\pm}$ moving in large scale magnetic fields, which are usually misaligned with the direction of pulsar's proper motion. They are distinct from the corresponding pulsar wind nebulae (PWNe), which are produced by e$^{\pm}$ trapped near the corresponding pulsars. Observations over the past decades show that filaments are common features of X-ray pulsars
\citep{2016A&A...591A..91P,2014ApJ...795L..27H,2014IJMPS..2860172P,2014A&A...562A.122P,2012ApJ...750L..39T,2016ApJ...828...70K,2021A&A...654A...4B}.

Filaments are usually asymmetric with the main filament much longer than the anti-filament, and there is no evidence of significant spectral variations, implying inefficient radiative energy loss of emitting e$^{\pm}$ \citep{2024ApJ...976....4D}. It is very challenging to explain these observations with diffusive transport of X-ray emitting TeV e$^{\pm}$ in large scale magnetic fields \citep{2014A&A...562A.122P}, since diffusion implies isotropy of emitting particle distribution and asymmetric injection of emitting particles into large scale magnetic fields is required \citep{10.1093/mnras/stz2819}.
On the other hand, these features suggest that the transport of emitting e$^{\pm}$ could be more or less ballistic with the scattering mean free path comparable to the filament length \citep{2021PhRvD.104l3017R, 2025ApJ...994..142H}. The X-ray emission process would be very inefficient, implying a very high injection power of X-ray emitting e$^{\pm}$. 

In this paper, we show that with a TeV e$^{\pm}$ injection power comparable to the spin down luminosity of the pulsar B2224+65 producing the famous Guitar PWN, its X-ray filaments can indeed be explained via ballistic transport of TeV e$^{\pm}$ in a large scale magnetic field with a slightly enhanced magnetic field of $\sim10\ \mu$G. In Section~\ref{sec:monte}, we first introduce 
a Monte Carlo code to simulate charged particle transport in a uniform magnetic field. It recovers the classical diffusion theory on long time scales \citep{1975Ap&SS..32...77F}. Its characteristics on short time scales are discussed.
In Section~\ref{sec:app}, 
we apply the model to the asymmetric X-ray filaments of pulsar B2224+65. Conclusions and discussions are given in Section \ref{sec:con}.

\section{Monte Carlo Simulation of Classical Diffusion in a Large Scale Magnetic Field}\label{sec:monte}

\begin{figure*}[htbp]

\epsscale{0.9}

\plottwo{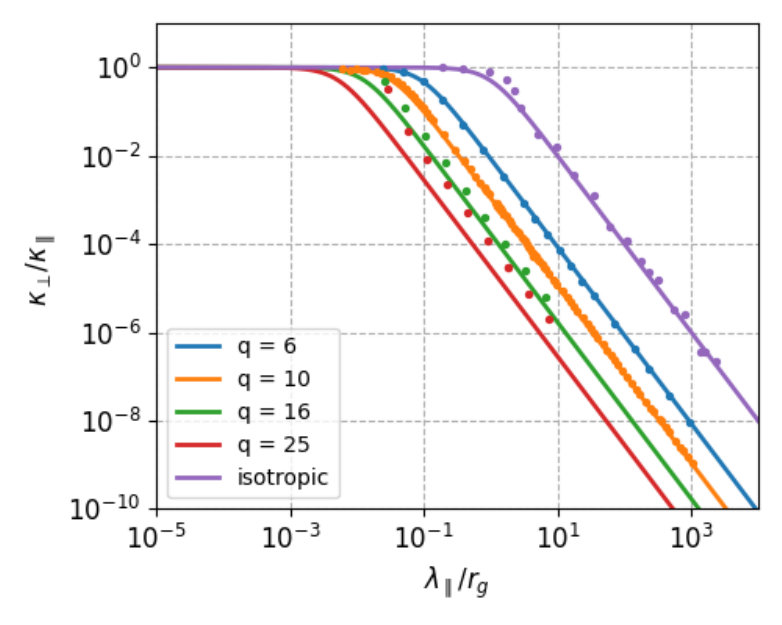}{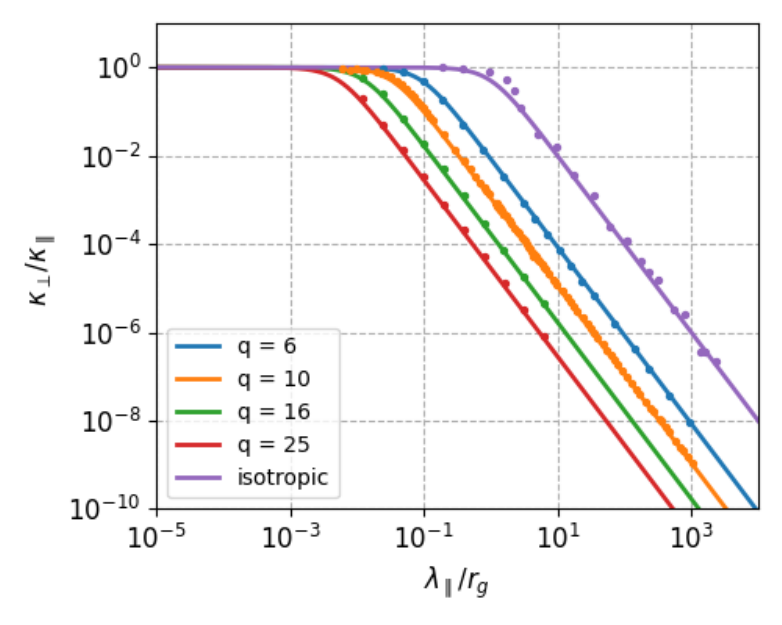}
\caption{Comparison of theoretical formulas with simulations. The purple curves correspond to isotropic scattering. The remaining four curves represent Gaussian scatterings with the standard deviation $\sigma=\pi/q$ with $q$ indicated. The left panel is for \(n = 100\) and the right for \(n = 600\). After sufficient scatterings with $n>q^2$, the numerical results agree to the theory.
\label{fig:1}}

\end{figure*}

Transport of high-energy charged particles in a turbulent magnetic field has been the subject of extensive investigations and has been approximated with diffusion. 
The classical theory \citep{1966ApJ...146..480J} does not consider the magnetic field wandering effects \citep{2008ApJ...673..942Y, 2014ApJ...784...38L} with the ratio of perpendicular diffusion coefficient $k_\perp$ to the parallel diffusion coefficient $k_\parallel$ given by  \citep{1970mtnu.book.....C,1965P&SS...13....9P,1975Ap&SS..32...77F}:
\begin{equation}
   \frac{\kappa_{\perp}}{\kappa_{\parallel}} = \frac{1}{1 + \left(\frac{\lambda_{\parallel}}{r_g}\right)^2}\,,
    \label{equ：equ1}
\end{equation}
where $\lambda_\parallel$ and $r_g$ represent the mean free path in the parallel direction and the Larmor radius of particles moving in perpendicular to the mean large scale magnetic field, respectively.
We verify this classical theory via Monte Carlo simulations by considering charged particle motion in a large-scale uniform magnetic field with
random changes of the particle velocity direction at a fixed time interval of $\tau$ (scattering time). Between two scatterings, the particle follows a helical trajectory. 

We first consider a uniform distribution of the particle velocity after each scattering (isotropic scattering).
For an instantaneous injection, there are therefore three parameters: the gyro-period $T=2\pi r_g/v$, the scattering time $\tau= \lambda_\parallel/v$, and the time $t$ since the injection that determines the number of scatterings $n=t/\tau$, where $v$ is the particle velocity and takes the value of the speed of light $c$ in this study. 
The ratio $T/\tau$ characterizes the turbulence intensity that determines how far the particle motion can deviate from the helical motion in a uniform magnetic field. 

For each values of $\lambda_\parallel/r_g$, we simulate 200,000 particles for a fix number of scattering $n$. We then calculate the standard deviations of the final particle location along and in perpendicular to the magnetic field to derive ${\kappa_{\perp}}/{\kappa_{\parallel}}$. The purple color in figure \ref{fig:1}
compares numerical results with equation \ref{equ：equ1}.

For the velocity change upon scatterings,
we also consider a Gaussian probability distribution of the scattering angle $\alpha$ between the particle's velocity before and after scattering:
\begin{equation}
    f(\alpha)={1\over\sqrt{2\pi}\sigma}\exp{\left(-{\alpha^2\over 2\sigma^2}\right)}\,,
\end{equation}
where $\sigma$ is the width of the Gaussian distribution.
and find an empirical diffusion coefficient formula similar to formula \ref{equ：equ1}:
\begin{equation}
\frac{k_{\perp}}{k_{\parallel}} = \frac{1}{1 + \delta \left( \frac{\lambda_{\parallel}}{r_g} \times \mathbf{q}^2 \right)^2}
    \label{equ：equ2}\,,
\end{equation}
where \(\delta\approx0.09\) and the standard deviation \(\sigma=\pi/q\). Figure \ref{fig:1} compare the simulation results with this formula for $100$ (left) and $600$ (right) scatterings. It is evident after sufficient scatterings with $n>q^2$, the simulation results agree well to the formula. In the following, we will only consider isotropic scatterings.

\begin{figure}[H]
\epsscale{1.0}
\plottwo{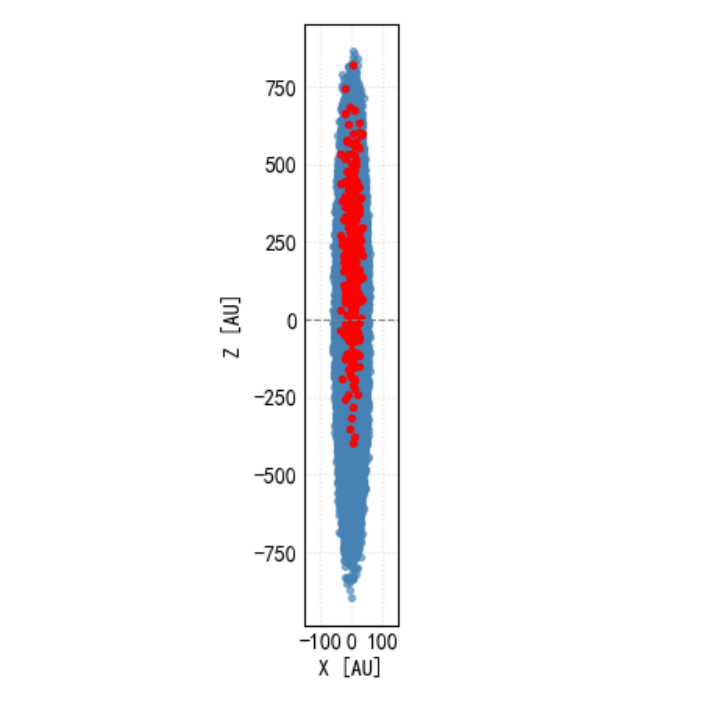}{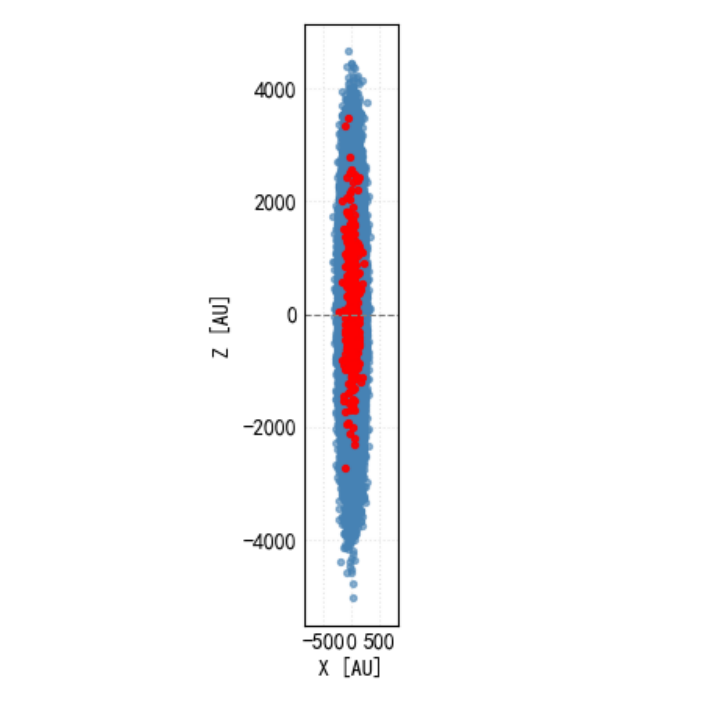}
\caption{Projected particle positions in the X–Z plane for $n = 4$ (left) and $n = 100$ (right). Blue dots are for all particles. Red dots mark those with $\theta\approx 30^\circ$. 
\label{fig:2}}
\end{figure}

\begin{figure}[htbp]  
\epsscale{1.0}
\plottwo{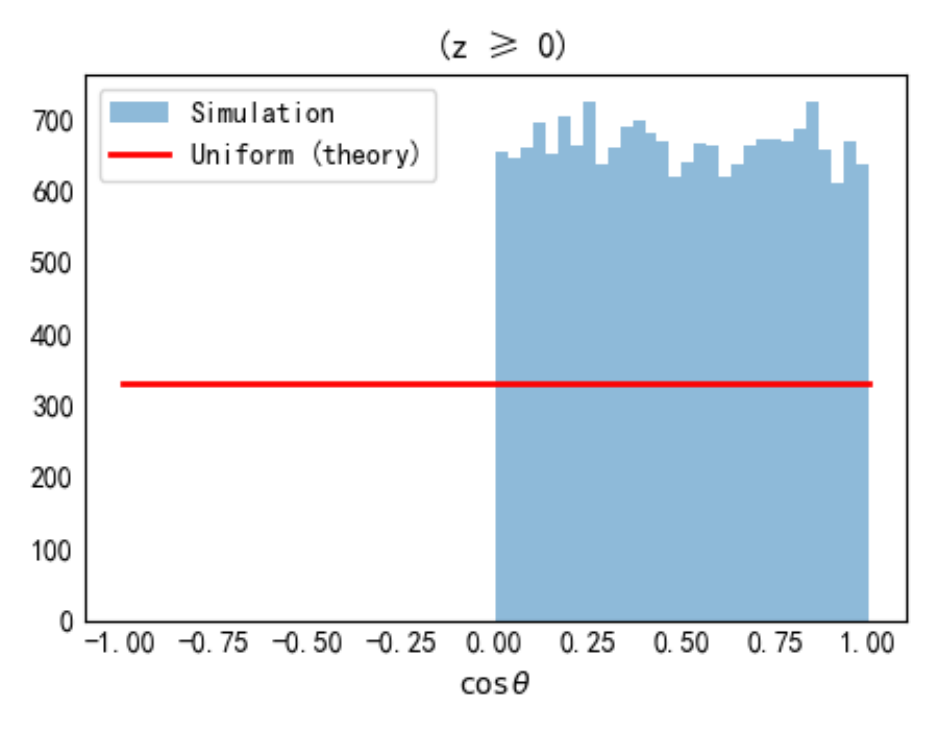}{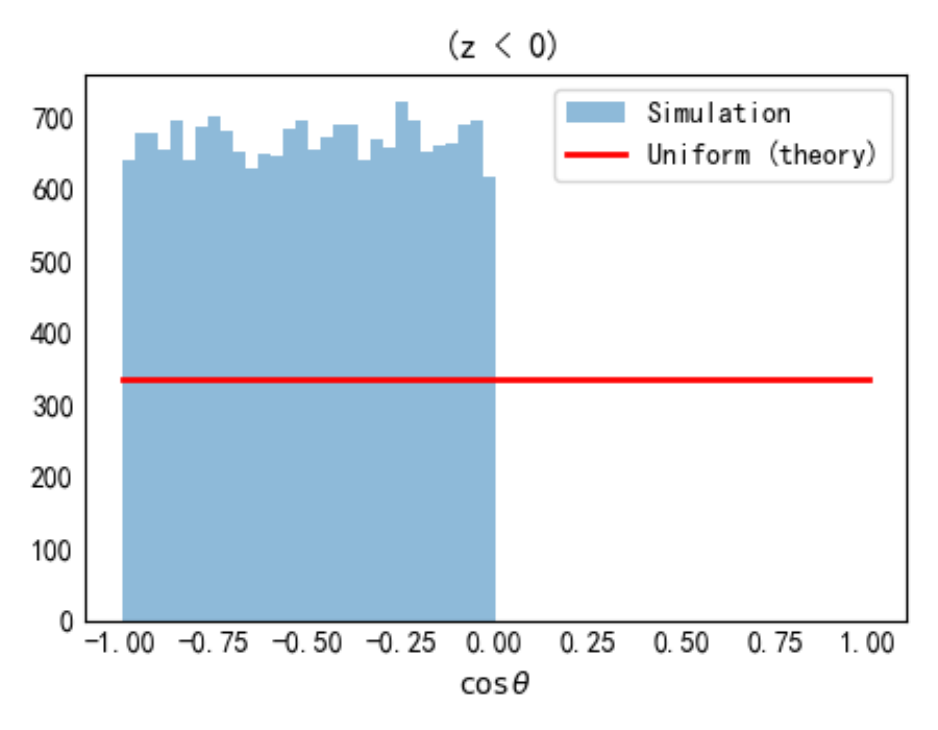}
\plottwo{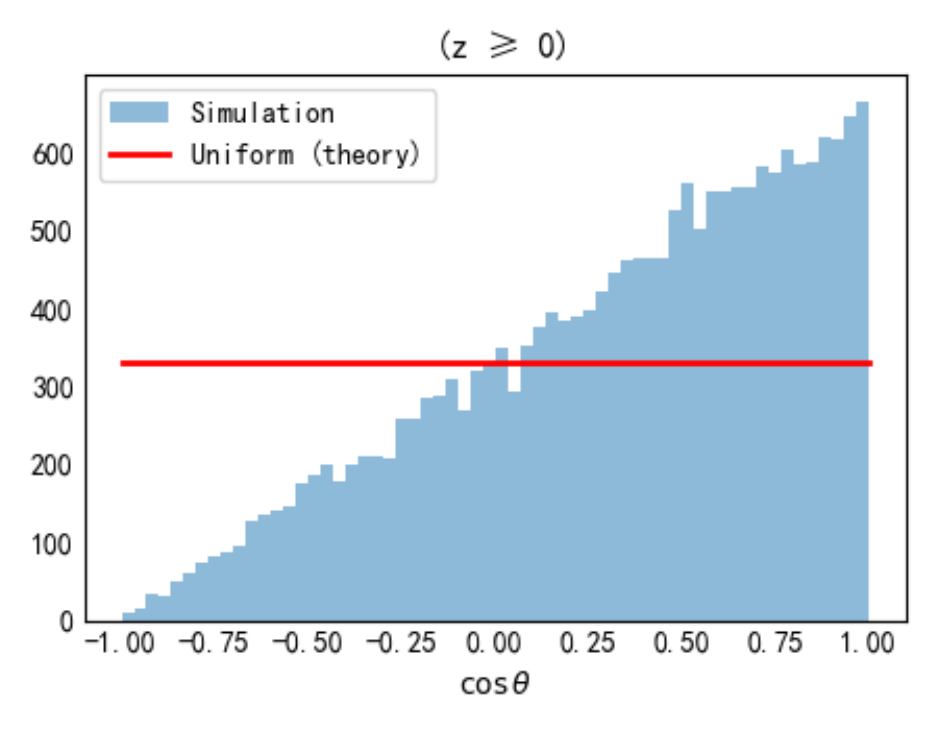}{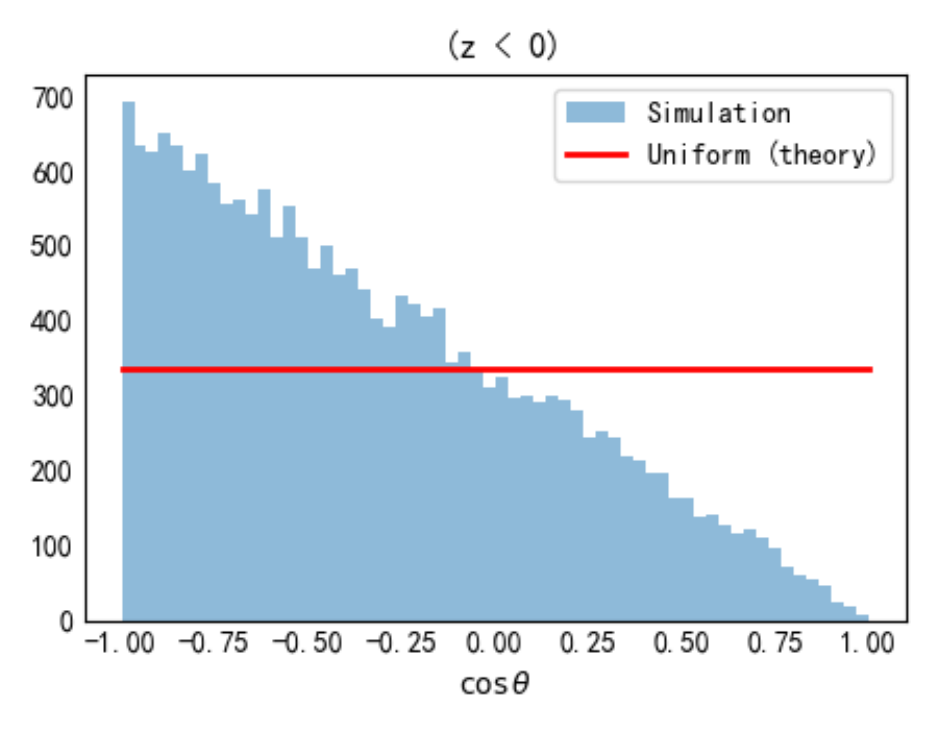}
\plottwo{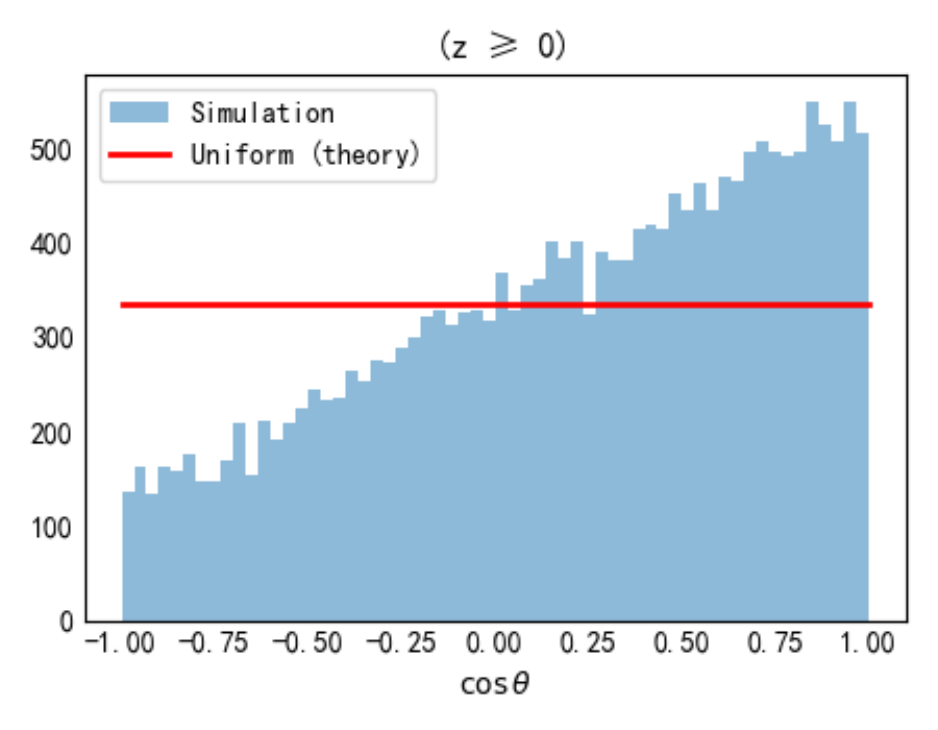}{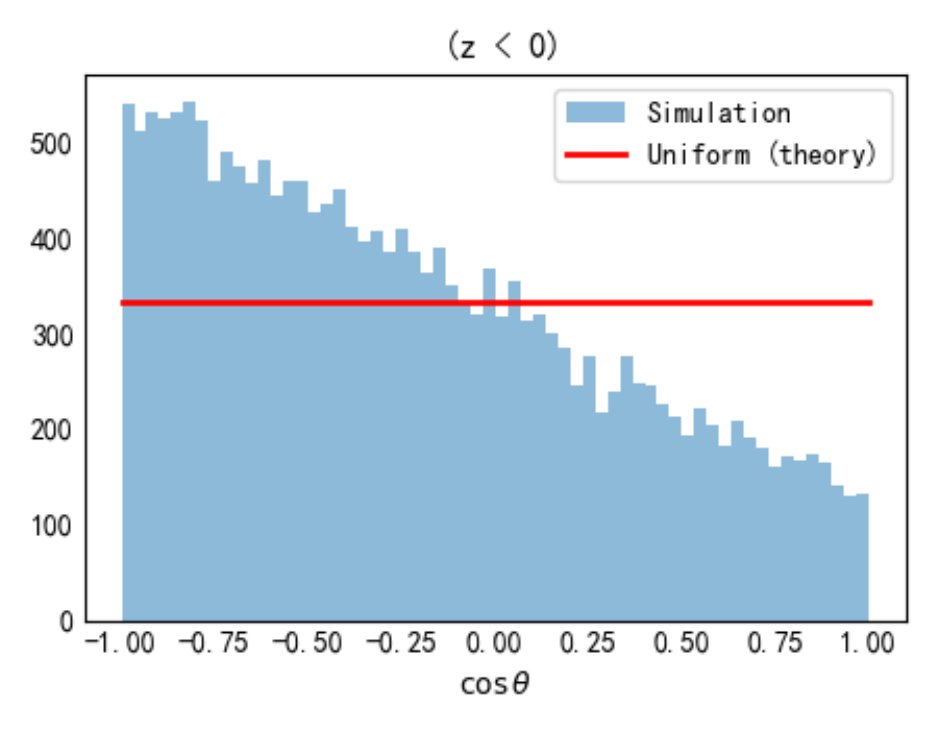}
\plottwo{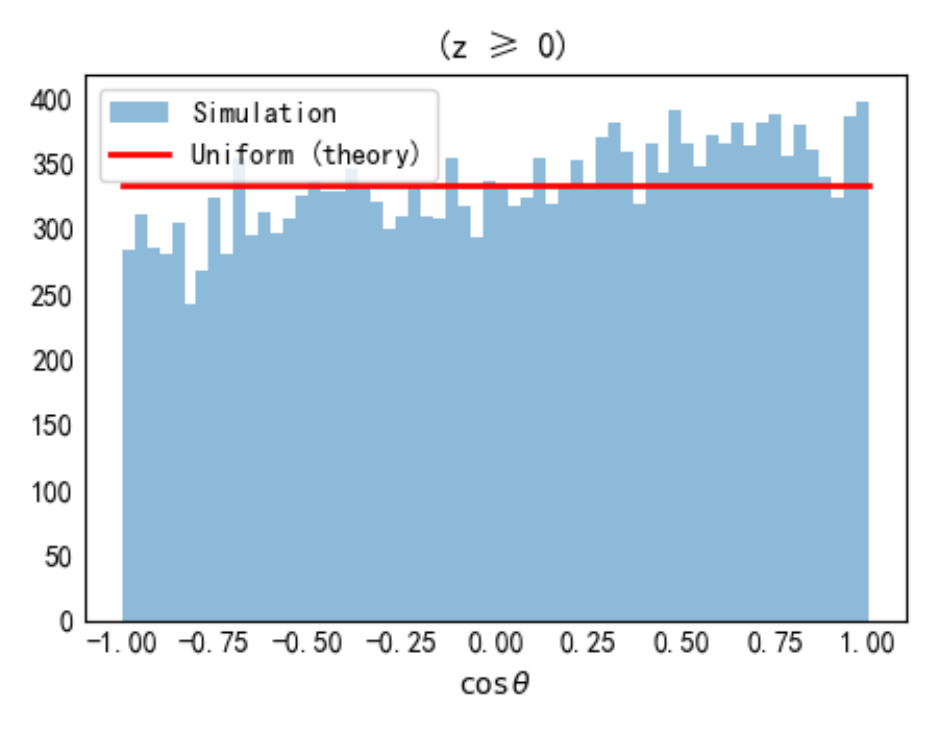}{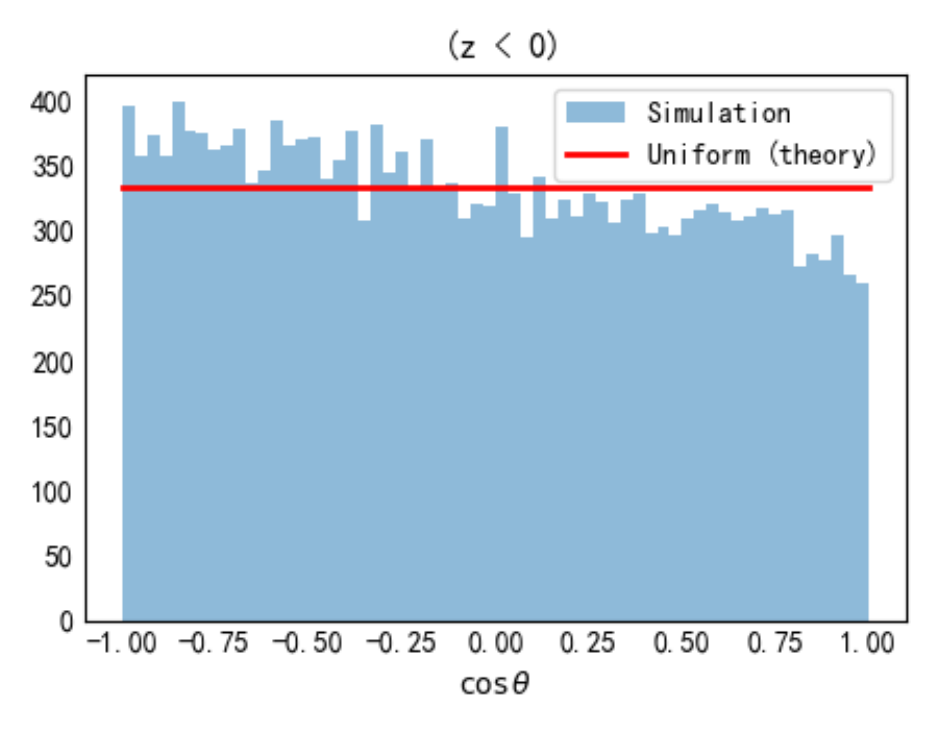}
\caption{Evolution of pitch angle distribution of all particles in the upper (left $Z\ge0$) and lower (right $Z<0$) regions. 
From top to bottom,
\(n\) = 0, 1, 4, 100, respectively.}
\label{fig:3}
\end{figure}

We are mostly interested in properties when the scattering is insufficient. We simulate 200,000 particles injected at the origin with the magnetic field in the $Z$ direction.
Since the system is axis-symmetric, we may focus on the pitch angle ($\theta$) distribution. Figure \ref{fig:2} shows the particle position projected in the $X-Z$ plane.
Figure \ref{fig:3} shows how the pitch angle distribution of all particles in the upper region (left \(Z \geq 0\)) and the lower region (right \(Z < 0\)) evolves with time. As \(n\) (propagation time) increases, each half-space approaches an isotropic distribution, but the two distributions  remain different even for $n=100$. 

\begin{figure}
\plottwo{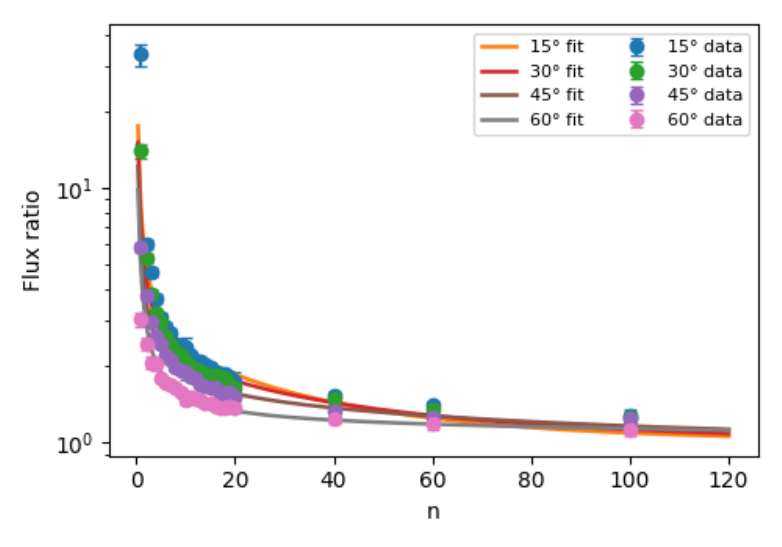}{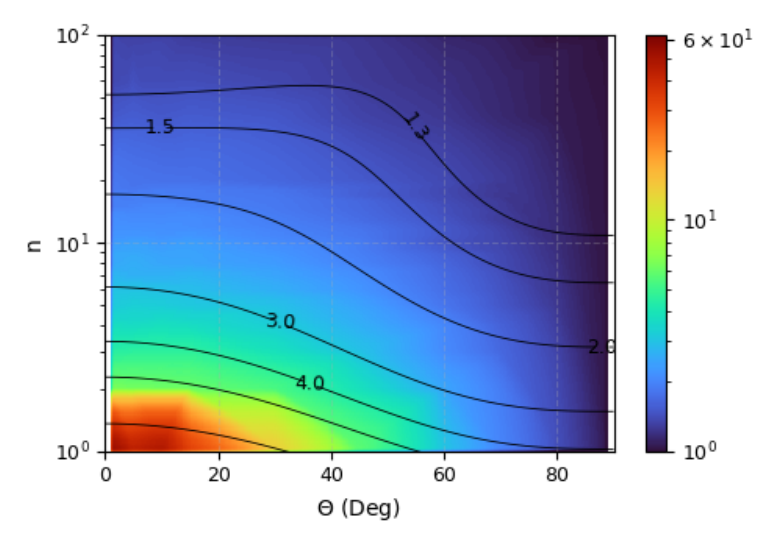}
\caption{Dependence of the flux ratio on the scattering times $n$ and the view angles $\Theta$ indicated on the figure. The lines are 
all for the empirical formula given by equation (\ref{equ：equ3}).
\label{fig:5}}
\end{figure}

Since radiation from relativistic particles are highly beamed along their velocity, if the Earth lies at angle $\Theta$ to the large-scale magnetic field (in the $Y-Z$ plane), only particles whose pitch angle is close to $\Theta$ are visible. 
Then the red dots in Figure \ref{fig:2} show that an apparent asymmetry  emerges even for an isotropic injection at the origin. 
One may use the ratio of numbers of particles with pitch angle $\Theta$ in the upper and lower regions to quantify this asymmetry. This ratio can be interpreted as the observed flux ratio between filament and anti-filament.
Figure \ref{fig:5} shows the dependence of this flux ratio on the scattering times $n$ and the view angle $\Theta$. As expected, the flux ratio drops with the increase of $n$ and $\Theta$ and approaches 1 for $\Theta=\pi/2$. 
We also find an empirical formula for the flux ratio
\begin{equation}
\left\{ 1 +A (\cos \Theta)^p \exp[-B  n  (\cos \Theta)^q] 
        \right\} \left[ 1 +\left( \frac{\varepsilon}{n} \right)^C \right]\,,
\label{equ：equ3}
\end{equation}
where \( A \approx 33.3 \), \( p \approx 1.60 \), \( B \approx 0.055 \), \( q \approx 1.88 \), \( \varepsilon \approx 0.97 \), \( C \approx 1.45 \). Figure \ref{fig:5} compares formula \ref{equ：equ3} with simulation results.



\section{Application}\label{sec:app}

\begin{figure}[ht]
\plottwo{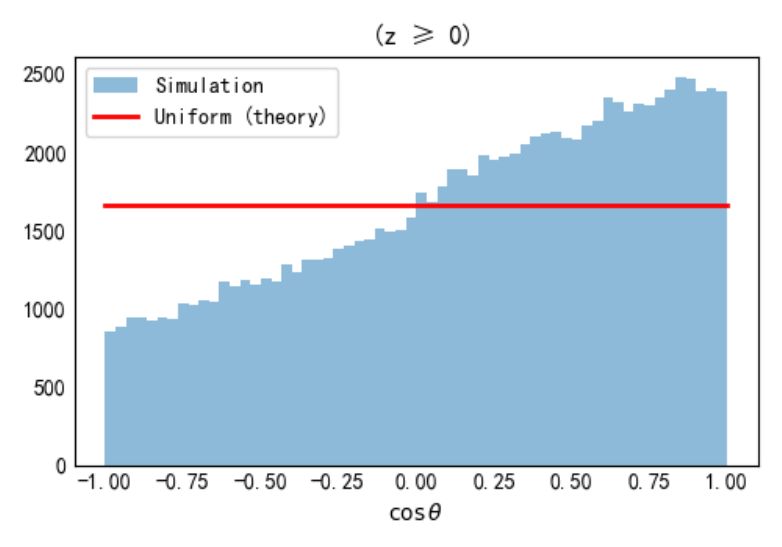}{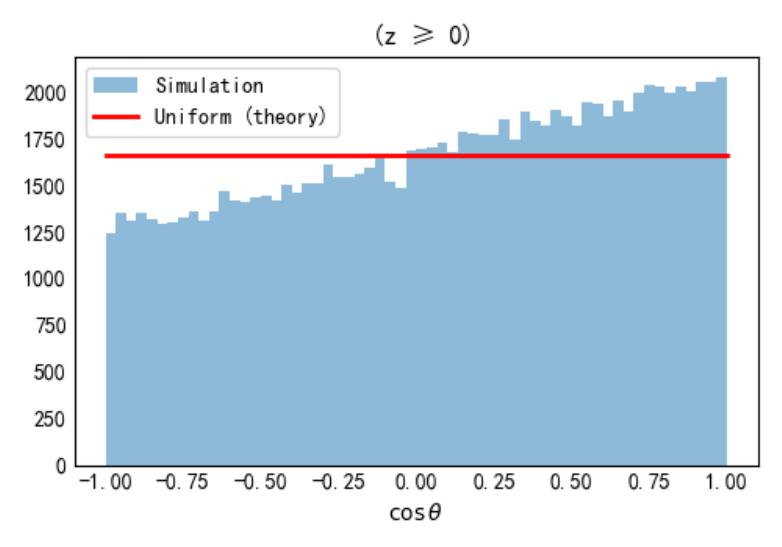}
\caption{The same as Fig. \ref{fig:3} but for stable continuous injection with $n = 20$ (left), and $100$ (right).
\label{fig:6}}
\end{figure}

To apply the model to observations of Guitar PWN (B2224+65), one needs to consider a continuous injection.
Figure \ref{fig:6} shows the pitch angle distribution in the upper region for a stable continuous injection. Note that the lower region has a similar distribution as the upper region (Fig. 
\ref{fig:3}).

The outflow of Guitar Nebula 
is formed via multiple injections from the central pulsar at different locations corresponding to different magnetic field lines \citep{2022ApJ...939...70D}.
We model this source with a stable injection at points moving in  perpendicular (in the X direction) to the large-scale magnetic field with a speed of 693 km s$^{-1}$, the measured value of the transverse proper speed. 
\begin{figure}  
\epsscale{0.8}
\plottwo{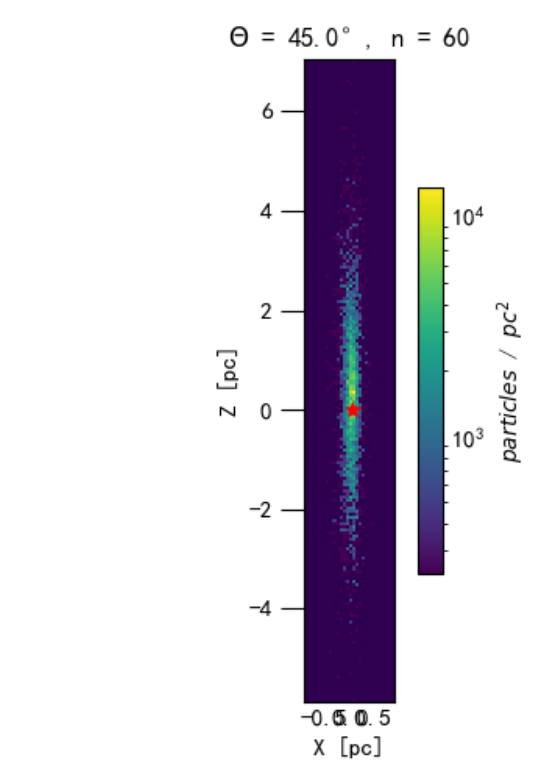}{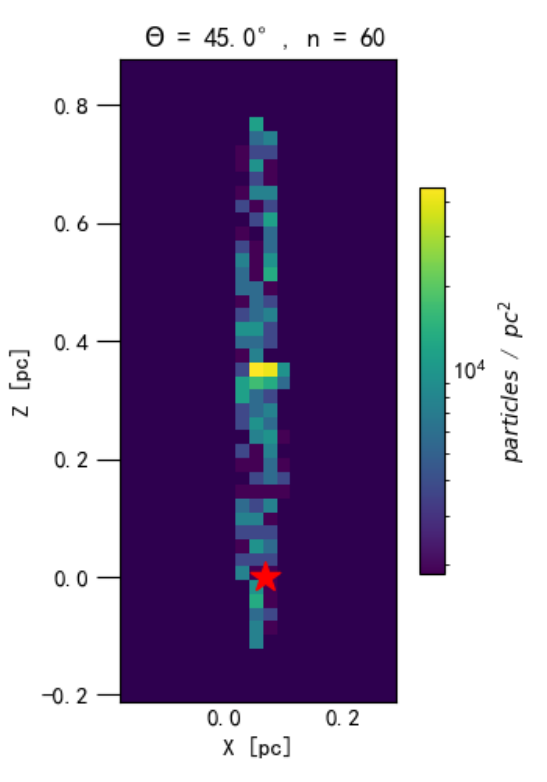}
\plottwo{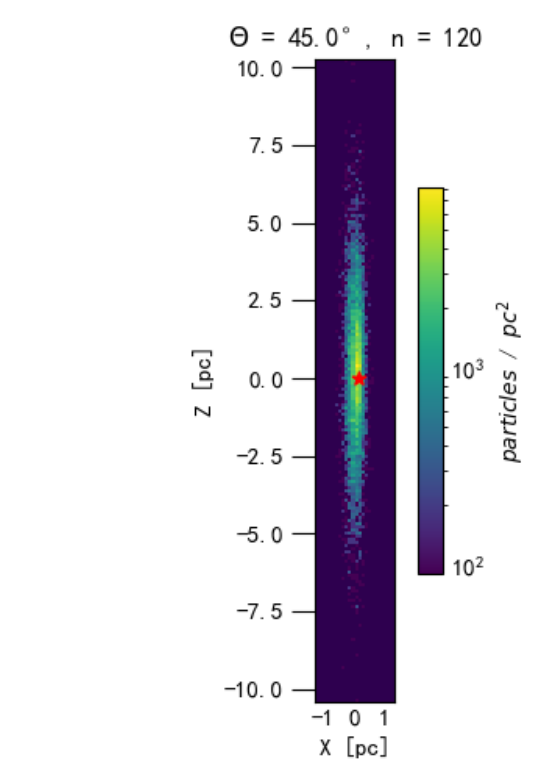}{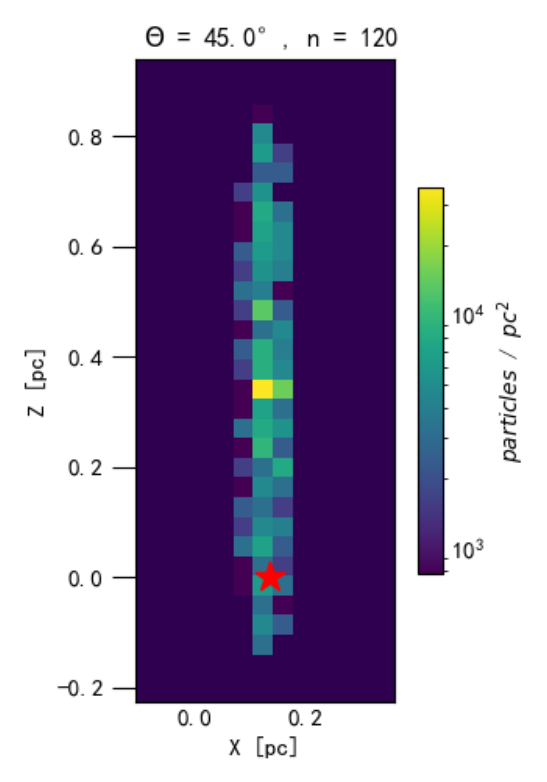}
\caption{Column density of particles with $\theta\simeq\Theta=45^\circ$ in the $X-Z$ plane for an injection times of 96 (upper) and 192 (lower) years. The right panels show those above a threshold of 5000 
pc$^{-2}$. The red star marks the location of the pulsar.}
\label{fig:7}
\end{figure}
Because the Guitar Nebula’s outflow spans only a few light-years while its cooling time exceeds several hundred years, radiative energy losses are neglected. 
Then the brightness of Guitar Nebula’s outflow should be proportional to the column density of particles whose pitch angle equals the viewing angle. 

The Guitar Nebula’s filament is 6–7 times longer than its anti-filament and carries about 11 times higher flux. The length-to-width ratio of the filament is about 13 \citep{2024ApJ...976....4D}. 
Although there is evidence that the injection rate of TeV e$^\pm$ into the X-ray filaments is not constant, the above features are consequences of hundreds of years of injection  \citep{2022ApJ...939...70D}. 
We therefore consider a steady-state solution near the pulsar with a time averaged injection rate. For high values of $n$, the diffusion approximation is valid. A steady-state near the site of injection can be achieved over a diffusion time scale $\sim R^2/D$, where $R$ is the size and $D$ is the diffusion coefficient. However, these particles will produce symmetric X-ray filaments. The observed asymmetry are results of ballistic transport of emitting particles \citep{2021PhRvD.104l3017R}. The steady-state is achieved on a smaller temporal and spatial scale. In X-ray observations, a nearby region is usually chosen as background \citep{2024ApJ...976....4D}. It is likely that the dimmer symmetric diffusion component, if exists on large scales, has been subtracted from the filaments as background. We therefore should focus on regions near the pulsar with high enough column density of emitting particles. 

According to observations, the distance to the Guitar PWN is $d=0.83$ kpc. The intrinsic length of the filament is $l_{fila}\approx0.6/\sin \Theta$pc.
Figure \ref{fig:7} shows a result that is compatible to the above mentioned features with $\Theta= 45^\circ$, $\tau= 1.6$ yr, $\tau/T = 4.6$. The left-hand panels show the column density along the $Y$ axis, and the right-hand panel shows the results above a threshold value of 5000 pc$^{-2}$. In principle, we should compute the line-of-sight integral to obtain the brightness profile. But for not too small viewing angle, the column density along the $Y$ axis is a good approximation. 
By comparing the upper and lower panels on the right-hand side, one can see that a steady-state near the pulsar is already achieved for $n=60$.
For each value of $\Theta$, by adjusting the scattering time-scale $\tau$, the scattering number $n$, the gyration period $T$, and the column density threshold, we can obtain similar results. The filament to anti-filament flux ratio is mostly determined by $\tau/T$, while the column density threshold determines the length. The length to width ratio is sensitive to the value of $T$. 
We introduce parameter $\beta$ to denote the flux ratio of the entire structure to its asymmetric part.When the asymmetric outflow reaches a steady state,
$\beta$ will increase linearly with $t=n\tau$ until radiative energy loss becomes important. Hence $\beta/t$ can be used to replace the column density threshold.

The characteristic energy of synchrotron radiation is given by $1.7\times10^{-11} B\gamma^2\sin\Theta$ keV, where $B$ is the magnetic field in Gauss and $\gamma$ is the Lorentz factor of emitting e$^{\pm}$. The radiative power of an emitting electron is given by 
$1.6\times10^{-15} B^2\sin^2\Theta\gamma^2$erg s$^{-1}$. The energy loss time is given by $5.0\times 10^{8} B^{-2}\gamma^{-1}\sin^{-2}\Theta$ yrs. The 0.5$-$7 keV unabsorbed flux is $F_{ob}=4.8\times10^{-14}$erg cm$^{-2}$ s$^{-1}$. The corresponding luminosity is 
$6.0\times 10^{30} $erg s$^{-1}$. Then the total number of emitting e$^{\pm}$ in the asymmetric filament is about $ 3.8\times10^{45}/B^2\sin^2\Theta\gamma^2$.
Combining this with the simulation data, the injection power of high-energy particles is readily calculated to be
\begin{align}
   P_{inj} &= 1.9\times10^{35}\left(\frac{14}{B/\mu {\rm G}}\right)^{2}\frac{1}{\sin^{2}\Theta}\notag\\
    &\quad \times\frac{\langle E \rangle/{\rm TeV}}{\langle E^{2}\rangle/(\rm TeV)^{2}}\frac{\beta}{t/{\rm yr}}\frac{0.9}{\eta}{\rm erg\:s}^{-1}
\label{equ：equ4}   
\end{align}

\noindent Here $\eta$ denotes the fraction of the total flux emitted in the observed band, and the angle brackets indicate the statistical average over the electron energy spectrum. The observed X-ray spectral index is 0.6, corresponding to an emitting electron spectral index of 2.2. We assume a broken power-law electron spectrum with a spectral index 1.2 (2.2) below and above a break energy less than 32 TeV derived from observations. With $B=28\ \mu G$ the injection power simplifies to 
$P_{inj}=1.3\times10^{33}\frac{\beta}{t/{\rm yr}}\times\frac{1}{\sin^{2}\Theta}{\rm erg\:s}^{-1}$.

\begin{figure}
\plotone{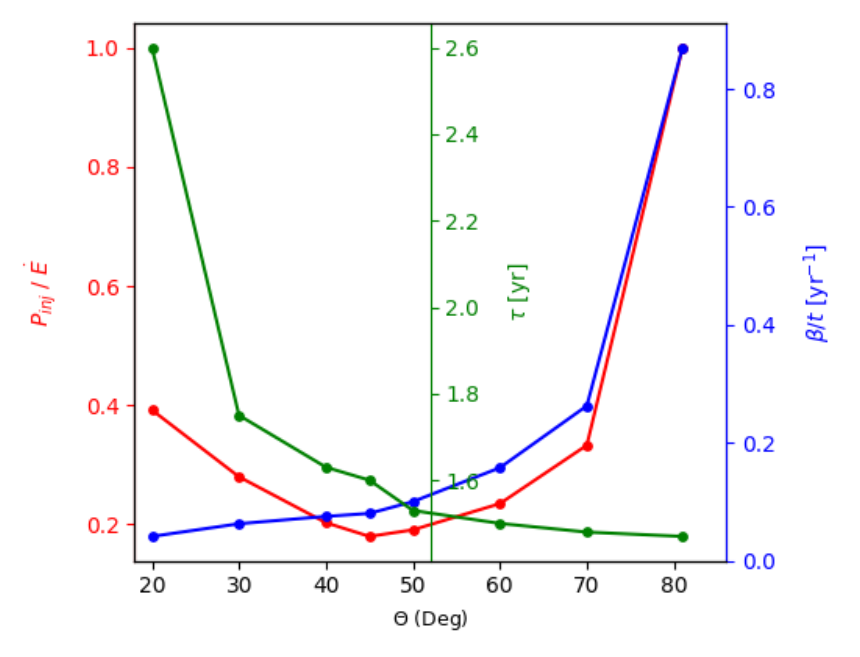}
\caption{Dependence of the injection power $P_{inj}$ (red), effective injection rate $\beta/t$ (blue), and scattering time scale $\tau$ (green) on the viewing angle $\Theta$ between the line of sight and the magnetic field. 
\label{fig:8}}
\end{figure}

The Guitar’s spin-down luminosity is 
$\dot{E}=1.2\times10^{33} {\rm erg\:s}^{-1}$. Assuming that the radiation is isotropic, the total radiative efficiency of the 
 outflow is $L\approx 0.6\%\dot{E}$. The cooling time-scale at this field strength is about $\tau_{\rm cool} \simeq 400$yr. Using our simulation results, we can obtain the specific value of the injection power. The red line in figure \ref{fig:8} shows its dependence on the viewing angle $\Theta$ and the corresponding $\tau$ (green) and $\beta/t$ (blue).  
 $t/(\beta\tau_{\rm cool})$ gives the radiative efficiency of the outflow. Using the minimum value of $\beta/t = 0.04/yr$ and $\tau_{cool}\approx400$ yr, the maximum radiative efficiency is about $6.3\%$. We have carefully computed the minimum time $t_{\min}$ required to reach the steady state for each $\Theta$: for $\Theta<20^{\circ}$ we obtain $t_{\min}>200\,\mathrm{yr}$, and for $\Theta\ge 40^{\circ}$ we find $t_{\min}\le 100\,\mathrm{yr}$. Consequently, energy losses must be considered for $\Theta<20^{\circ}$, and our present model is valid only for $\Theta>20^{\circ}$. $\tau$ scales with the projection factor $\frac{1}{\sin\Theta}$: at small $\Theta$ the factor is large, so $\tau$ is large, while for $\Theta\ge 50^{\circ}$ the factor changes very little and $\tau$ becomes practically constant. 

 The minimum of $P_{inj}/\dot{E}$ is about $18\%\dot{E}$ and occurs near $\Theta=45^\circ$. At $\Theta=80^\circ$ the injection power rises to $\dot{E}$. For $\Theta<60^\circ$, $\beta/t$ varies slowly with $\Theta$, so $P_{inj}$ is affected mainly by $1/\sin^{2}\Theta$, which changes rapidly below $\Theta=45^\circ$.
 Once $\Theta$ exceeds $60^\circ$, 
 $\beta/t$ rises steeply to produce the observed asymmetry while $1/\sin^{2}\Theta$ flattens, forcing $P_{inj}$ to rise rapidly again. 
 The electron-injection power 
 is compatible to values derived from other studied \citep{2018ApJ...864L...2T,2014JHEAp...1...31T}. We also study the effect of small pitch angle distribution of injected particles on the total injection power. Small pitch angle distribution can result from magnetic momentum conservation if particles propagate from a strong field into a weak field \citep{2024NatAs...8.1284K}.
 For $\Theta =45^\circ$, the injection power can be reduced to $44\%$ of its value for the isotropic injection. 

\section{Discussion and Conclusions}
\label{sec:con}

In this letter, we present a simple model for asymmetric X-ray filaments from pulsar winds. Its application to Guitar PWN shows that with a magnetic field of tens of $\mu$G, an injection power of a few tens of percent of the spin down luminosity, and a scattering mean free path comparable to the filament length, the observed characteristics of the filament can be reproduced. The model predicts a symmetric diffuse background which can be tested with observations with a large FOV and a high sensitivity. The model has not considered large scale structure of the magnetic field, which may explain the helical structure seen in lighthouse PWN \citep{2012ApJ...750L..39T, 2016A&A...591A..91P, 2014A&A...562A.122P}. 
Magnetic field wandering effects are also important on large scales \citep{2014A&A...562A.122P, 2019PhRvL.123v1103L} and should be considered when probing this diffuse component. The above results show that such X-ray observations can be used to probe properties of the interstellar magnetic fields and may explain TeV halos of some powerful pulsars \citep{2017Sci...358..911A, 2021PhRvL.126x1103A}.
The model does not consider radiative energy loss that may play an important role in filaments observed in TeV \citep{author2025Peanut-shaped}. 

\begin{acknowledgments}
We thank Prof. Lazarian for helpful discussion. This work is supported by the National Natural Science Foundation of China under the grants 12375103 and U1931204.

\end{acknowledgments}

\bibliography{export-bibtex}{}
\bibliographystyle{aasjournalv7}

\end{document}